\documentclass[aps,pra,superscriptaddress,twocolumn]{revtex4-1}

\usepackage{mathrsfs}
\usepackage{amsmath}
\usepackage{amssymb}
\usepackage{amstext}
\usepackage{graphicx, epsf}
\usepackage{enumerate}
\usepackage{bbm}
\usepackage[english]{babel}

\newcommand{\ignore}[1]{}

\newcommand{\be}{\begin{equation}}
\newcommand{\ee}{\end{equation}}

\begin{document}

\title{Topological order away from equilibrium}

%\classification{03.65.Vf, 03.65.Ud}
%\keywords      {Statistical Mechanics, Topological Order, Quantum Quench, Entanglement}

\author{Alioscia Hamma}
\affiliation{Center for Quantum Information\\ Institute for
  Interdisciplinary Information Sciences\\ Tsinghua University\\
   Beijing
  100084, P.R. China}

\begin{abstract}
Topological ordered states are exotic quantum states of matter that defy the usual description in terms of symmetry breaking and local order parameters. The type or order they feature is of non-local, topological nature, and it allows such systems to present unusual and interesting behaviours, like robust degeneracy, and anyonic excitations. Here we focus on the behaviour of such systems away from equilibrium. After a sudden change of the Hamiltonian - the so called quantum quench - quantum many body systems show thermalization, unless the quench is integrable. We show how in topologically ordered systems  out of equilibrium behaviour do not follow the usual thermalization behaviour.
\end{abstract}

\maketitle

%%%%%%%%%%%%%%%%%%%%%%%%%%%%%%%%%%%%%%%%%%%%
%% MAINMATTER
%%%%%%%%%%%%%%%%%%%%%%%%%%%%%%%%%%%%%%%%%%%%

\section{Introduction}
A scientific theory is the more successful the largest is its explanatory power. 
For this reason, one of the greatest scientific achievements of theoretical condensed matter is the Landau theory of symmetry breaking and local order parameters. Until recently, indeed, all states of the matter could be classified about their symmetry breaking pattern\cite{Goldenfeld}. Nevertheless, often great progress in science is obtained by finding very exotic and rare situations, because they defy the previous paradigm, and allow to think in a completely different way. Sometimes, this allows for a newer whole general theory. There is no better example perhaps than the weird results about double slits experiments, which opened up the new spooky world of quantum mechanics. 

In this paper, we are interested in some exotic states of matter that are characterised by some very elusive property: topology. Topology is exotic because usually nature likes to be topologically trivial. For example, nature seems never invented the wheel! Do you know organisms with wheels? This is because, in order to "grow", an organism needs to transport biological material from one part of its structure to another, and this forbids to have parts freely revolving around another. So here we have a law of biology: no wheels! This is a good law because it is true in so many instances. But again, the exotic cases in which the law may be defied, are of greatest interest for the scientist. When biologists discovered the flagellum, that is used to propel some microorganisms, they found the wheel! So nature is not always topologically trivial. Does something similar happen also in materials? 

Until about thirty years ago, it would seem that there was nothing topologically interesting in materials. They were all described by symmetry breaking, and phases could be detected by doing local measurements of the so called local order parameter. Then, in 1982, Tsui, Stormer and Gossard \cite{tsui} discovered the fractional Quantum Hall Effect (FQHE). Such liquids can exist in different phases without any symmetry change. Therefore Landau's theory is not complete. A similar behaviour was discovered in lattice gauge theories, where different phases, like the pure gauge and the paramagnet both have no symmetry breaking, and they are separated by a quantum phase transition\cite{wegner}.

It is now clear that these states of the matter are distinguished by topological properties of the ground state wave-function. For instance, in the case of the FQH liquids, these topological properties dictate the dancing rules for the highly constrained motion of the electrons in the two dimensional electron gas\cite{Wen-1}. 

More recently, topological phases have become very important for quantum information processing. This is exactly because of their topological character. The main obstacle that has to be overcome in order to build a quantum computer is decoherence. Decoherence is a nasty enemy but it is a local process. There is perhaps hope that topological features can be resilient under decoherence, and interesting enough to host quantum computation\cite{kitaev}.
%\begin{figure}
  %\includegraphics[height=.26\textheight]{flagellum.jpeg}
  %\caption{Bacterial flagellum - Figure courtesy of Wikipedia. The propulsion mechanism consists in a flexible tail freely revolving around a hub. There are wheels -non trivial topologically- in biology.}
%\end{figure}

In this paper, we show the behaviour of the simplest example of topologically ordered state away from equilibrium. We study the case of the so called \textit{Quantum Quench} (QQ). The system is prepared in the ground state of a certain Hamiltonian $H_0$ at the time $t=0$ and then the Hamiltonian is suddenly switched to $H_1$, usually by changing some external field. The system is then not in an eigenstate of $H_1$ and will undergo unitary time evolution. This kind of behaviour is now experimentally investigable in systems of ultracold atom gases \cite{bloch}. Typically, it is found that such quantum systems thermalize. This a beautiful topic to investigate the foundations of statistical mechanics\cite{nature}. What we are concerned about though, is the behaviour of topological order under a QQ. If these states need to be robust against general decoherence, they must conserve their topological order also under a QQ. 
In order to study the behaviour of Topological Order away from equilibrium, we consider the Kitaev's toric code\cite{kitaev}, which is the simplest example of topologically ordered system.

\section{Kitaev's toric code}
\begin{figure}\label{lattice}
  \includegraphics[height=.18\textheight]{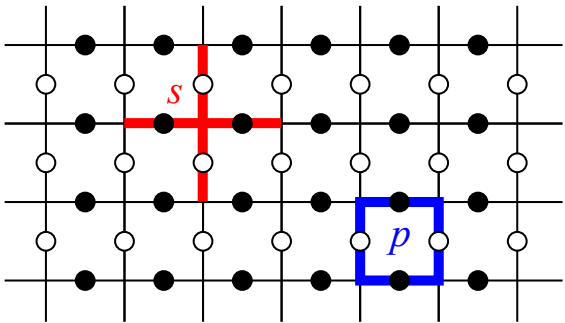}
  \caption{Kitaev's toric code.  The
physical spins are on the horizontal (black circles) and the vertical
(white circles) bonds of the square lattice. The star operator (red) is a four body operator involving the four spins extruding from a site $s$, while the plaquette operator (blue) involves the four spins around a plaquette $p$.}
\end{figure}

The Kitaev's toric code consists of a square lattice with periodic boundary conditions, that is, a torus, on whose bonds are placed spins one-half. On a square lattice of $L\times L$ sites there are therefore $2L^2$ bonds. The total dimension of the Hilbert space is thus $2^{2L^2}$. In a lattice gauge theory, the gauge invariant Hilbert space is obtained by projecting from the total Hilbert space, to a subspace where some local constraints must be obtained. Consider the the plaquette $p$ (See Fig.\ref{lattice}). There are $L^2$ plaquettes on the lattice. Any operator $B_p$ with support on the spins in $p$ is a local operator. We may ask as constraint that the allowed wave functions $\psi$ have to obey the law $B_p\psi=\psi$. This is analogous with the Gauss' law in electrodynamics. Now, we do not really want to restrict the Hilbert space, but if we place the term $H_B(U) = -U\sum_p B_p$ in the Hamiltonian, and if $U$ is a very large energy scale, at low energies (much smaller than $U$) the states must obey the aforementioned constraint. We can then say that we have the gauge theory emerge at low energy. In this theory, we choose $B_p = \prod_{j\in p} \sigma^z_i$. In other words, it is the flux of the $Z_2$ field through the plaquette $p$. The operator $B_p$ has two eigenvalues $\pm 1$, and for this reason the resulting low energy theory is a $Z_2$ lattice gauge theory. At this point, all we are saying is that the allowed states at low energy must respect this constraint. Notice that this constraint is obeyed for every spin configuration that consists of spins flipped down around any loop we can draw on the lattice. We then say that our low-energy theory only contains loops. 

In order to have a gauge theory, the full Hamiltonian must be made of terms that commute with the gauge constraint $B_p=1$. We can easily see that the right Hamiltonian is 
\begin{equation}
H_{} = H_B(U) - J \sum_s \prod_{j\in s} \sigma_j^x -  \sum_l h_l \sigma^z_l
\end{equation}
where both $J,t\ll U$. 
The ground state of this Hamiltonian is the same of the $Z_2$ gauge theory, regardless of the value of $U$. At high energy, this Hamiltonian does not describe the $Z_2$ theory any longer. The Hamiltonian with $h_l=0$ is the so called Kitaev's toric code $H_{TC}$. We say that the ground state of $H_{TC}$ is {\em topologically ordered}. What does it mean? First of all, we notice that there is a gap. The ground state is obtained by setting all the $A_s\equiv  \sum_s \prod_{j\in s} \sigma_j^x$ to $+1$, and the ground state energy is thus $E_0=-L^2(U+J)$. Every time we flip a $A_s$ from $+1$ to $-1$, we create an excitation (localized at $s$) of energy $E_0+2J$. Also notice that, because of the periodic boundary conditions, the product of all the $A_s$ is nailed to $+1$. As a consequence, we can only flip an even number of $A_s$. So the first excited state is the degenerate state obtained by flipping any two $A_s, A_{s'}$ at two sites $s,s'$ and has energy $E_1=E_0+4J$. The degeneracy comes from the fact that $s,s'$ can be located anywhere in the lattice. The gap $E_1-E_0=4J$. Now, we wonder whether the ground state is unique or not. Since $A_s=+1$ and $B_p=+1$ are local constraints, we do not expect that they can be spontaneously broken. This is the content of the Elitzur theorem: {\em a local symmetry cannot be spontaneously broken}\cite{elitzur}. We know that symmetry breaking provokes a degenerate ground state of different symmetry broken sectors. So we know one thing for sure, if there is degeneracy, this is not due to symmetry breaking. But is this possible at all? Consider a ground state for $H_{TC}$, namely, $\psi_0$. Now consider the operator that flips all the spins around the torus in one of the two directions, call it, $W^x_1$. It is easy to see that $\left[H_{TC}, W^x_1\right]=0$ and that $\langle \psi_0|W^x_1|\psi_0\rangle =0$. This means that the state $|\psi_1\rangle \equiv W^x_1|\psi_0\rangle$ is (1) orthogonal to $\psi_0$ and, (2) has the same energy. Similarly, we can construct another orthogonal degenerate state by flipping spins in the other torus direction with $W^x_2$ and in this way we can construct four orthogonal states $\psi_0, W_1^x\psi_0, W^x_2\psi_0,W^x_1 W^x_2\psi_0$. These four state span a four dimensional degenerate subspace that is the ground state manifold of $H_{TC}$ without any symmetry being spontaneously broken! 

It is very easy, in this setting, to prove the absence of spontaneous symmetry breaking, that is, Elitzur's theorem. Imagine there is some operator that has to {\em spontaneously} take a definite value in the ground state. Spontaneously means that it must be some operator that we can send to zero and look for a residual value. Without loss of generality, consider the operator $\sigma_j^z$. Then we want to compute $M_j=\langle \psi_0|\sigma^z_j|\psi_0\rangle $. We can choose a site $s$ such that $j$ is one of the spins that comes out of it, $j\in s$. Also, since $A_s|\psi_0\rangle =|\psi_0\rangle$, we can write $M_j=\langle \psi_0|A^\dagger_s\sigma^z_j A_s|\psi_0\rangle $.  Now, notice that, because $j\in s$, then $\{ A_s, \sigma^z_j\}=0$ and therefore we have $M_j=\langle \psi_0|\sigma^z_j |\psi_0\rangle = -\langle \psi_0|A^\dagger_s A_s \sigma^z_j \psi_0\rangle =0$. So no local operator can take different values in the four ground states. Of course, the operators $W^x_1,W^x_2$ do take different values in the four ground states, but they are non local! And so there is no spontaneous symmetry breaking between them. A consequence of this is the following, let $I$ be a finite set of contiguous spins, and consider the  density matrix of the ground state reduced to $I$ obtained by tracing out all the degrees of freedom outside $I$. This reduced density matrix represents the outcome of all the possible local (to $I$) measurements on the system.  It turns out that since no local measurement can distinguish the four states, then their four local density matrices must be identical. So this is a working definition of topological order, a gapped system, with a degenerate ground state, whose degenerate ground states are locally indistinguishable.

We have seen that at zero field, the ground state must be made of superposition of loop configurations. Indeed, since they are all equivalent, and since the $A_s$ term likes to make loops fluctuate, energy is minimized if loops of all sizes are equally represented. Indeed, since $A_s=1$, if we call $G$ the set of all the possible products of some $A_s$ operators (which makes a group), then the ground state $\psi_0$ can be written as $|\psi_0\rangle =|G|^{-1/2}\sum_{g\in G}|g\rangle$, where $g$ is the operator obtained by taking the product of all the $A_s$ operators with $s$ in some set $S$, which flips all the spins on the state with all spins up: $g= \prod_{s\in S} A_s$, and $|g\rangle = g|\uparrow\ldots\uparrow\rangle$. Then it is easy to see that $A_s\psi_0 =B_p\psi_0=\psi_0$ and that therefore $\psi_0$ is indeed a ground state. Now we can see how different ground states are locally the same. The key element is that, given $I$ a finite region of spins (and to be finite, it must also be contractible, i.e., not go around the torus), there is a way to deform $W_1^x\mapsto \tilde{g}W^x_1$ so that its support has no intersection with $I$, where $\tilde{g}$ is in $G$, i.e., is a product of $A_s$ operators. Then, using the fact that $\tilde{g}\psi_0=\psi_0$, we have
\begin{eqnarray}
\nonumber
\rho_I^{(1)} &=&\mbox{tr}_{\bar{I}}|\psi_1\rangle\langle\psi_1| = \mbox{tr}_{\bar{I}} W^x_1|\psi_0\rangle\langle\psi_0|W^x_1=\\
\nonumber
&=& \mbox{tr}_{\bar{I}} \tilde{g}W^x_1|\psi_0\rangle\langle\psi_0|W^x_1\tilde{g} =\\
\nonumber 
&=&
\sum_{g,g'\in G}  \tilde{g}|g_I\rangle\langle g_I' |\tilde{g}\langle g_{\bar{I}}'|(W^x_1)^\dagger W^x_1|g_{\bar{I}} \rangle\\
\nonumber
&=& \sum_{g,g'\in G}  \tilde{g}|g_I\rangle\langle g_I' |\tilde{g}\langle g_{\bar{I}}'|g_{\bar{I}} \rangle\\
=\rho_I^{(0)} 
\end{eqnarray}

where $\bar{I}$ is the complement of $I$ on the lattice.

As we switch on the external field $h$, or any other perturbation, this property is stable for a while. The external field acts like a string tension, so, as the field strength increases, the amplitudes for larger strings become smaller. At a critical value of the field strength,  the gap closes and a quantum phase transition occurs. So there is a whole gapped phase in which this property is maintained, and that is why topological order is a property of a {\em phase}.  \cite{HammaLidar:06, Trebst:06, spyr}. Notice that a rich phase diagram can be obtained even if we break all gauge symmetries, like introducing fields in different directions\cite{vidal}. Topological order does not rely on any symmetry, not even the gauge one. 
%%%%%%%%%%%%%%%%%%%%%%%%%%%%%%%%%%%%%%%%%%%%
%% Sample figure:
%%
%% The option [height=...] scales the picture to the given height,
%% without it it would be printed at its nominal size
%%%%%%%%%%%%%%%%%%%%%%%%%%%%%%%%%%%%%%%%%%%%

\section{Topological order is a pattern of entanglement}
The fact that local measurement cannot distinguish the different topological states in the ground state, is at the root of the hope that such systems can host robust quantum computation at the physical level. In particular, we can encode a qubit in the degenerate ground space, and ope that the quantum memory will be stable\cite{dennis}. However, this property is very elusive, after all, we cannot find it out from local measurements. Moreover, it is not clear {\em why} this property is there in the first place. What is in the ground state wave function that gives rise to such a feature? We would like to show that topological order is the property of a wave-function, regardless of Hamiltonians, excitations, and other states. There are solid reasons to do this. One, of very practical importance, is that, if topological order is the property of a wave-function, then we can talk about this property also when the system is away from equilibrium. And, for the reasons explained above, this is something very important. 
\begin{figure}\label{Figure2}
  \includegraphics[height=.11\textheight]{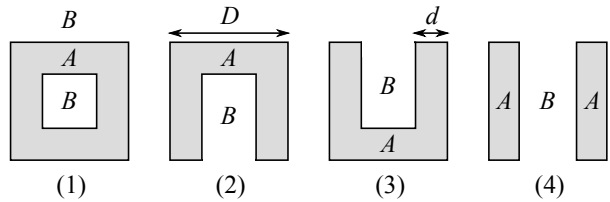}
  \caption{The four subsystems used to define topological R\'enyi entropy as $2S_\alpha^{top}= S_\alpha^{(1)}+S_\alpha^{(3)}-S_\alpha^{(2)}-S_\alpha^{(4)}$}
\end{figure}

If we are looking for something that is very specific of a quantum mechanical wave-function, then in some sense that must be related to entanglement. Entanglement is indeed what makes a many-body quantum wave function something more than a classical distribution of probability. Computing the entanglement entropy requires finding the eigenvalues of the reduced density matrix to a subsystem $I$. As it was shown in \cite{te1}, the reduced density matrix is proportional to a projector, and therefore its entanglement spectrum is flat. In particular, we have 
\begin{equation}
(\rho_I^{(0)})^2 = \frac{|G_A\times G_B|}{|G|}\rho_I^{(0)} = 2^{|\partial I|-1}\rho_I^{(0)}
\end{equation}
where $|\partial I|$ is the length of the boundary of the region $I$. 
Of course a flat spectrum means that also all the R\'enyi entropies are the same\cite{renyi}. So, regardless of the R\'enyi index alpha, we have 
\begin{equation}
S_\alpha(\rho_I^{(0)})= |\partial I|-1
\end{equation}
The $-1$ is an $O(1)$ correction to area law. It is also the logarithm of the dimension of the gauge group (or the quantum dimension, if one thinks of the excitations). The prefactor of the area law cannot be universal, but what abou the $O(1)$ correction? Maybe this quantity can serve to detect topological order, as it was proposed in \cite{te1,te2,te3}. One way to interpret this correction, is to think of it as a constraint on the configurations allowed on the boundary of $I$\cite{renyi}. A  convenient way to extract this quantity, is to define the subregion $I$ in four different fashions, and then take a suitable linear combination of the entropies \cite{te2,te3}, see  Fig.\ref{Figure2}.

At this point, we must ask ourselves if this quantity is really the marking of the topological phase. It turns out that the answer is positive, indeed, for both $\alpha=1$ and $\alpha=2$ it has been shown, with numerical and analytical methods\cite{num, gab1}, that is quantity is constant within the whole topological phase, and then drops suddenly to zero after the quantum phase transition. We can conclude this section by stating that topological order is a pattern of entanglement in the wave function, that can be revealed in topological constraints on the boundary between subsystems, resulting in robust corrections to area law within the phase. 
This kind of entanglement is non-trivial. Most states of the matter possess trivial entanglement. Indeed, there is a way to analytically deform such states by adiabatic evolution with some local Hamiltonian, that would completely disentangle the state. Of course such Hamiltonian must break the symmetries of the original Hamltonian or a quantum phase transition would occur, thereby breaking adiabaticity. But this is not the point. What matters here is that, if we disregard symmetry, usual states can be disentangled in a completely factorized state. In this sense their entanglement is trivial. One may suspect that {\em all} the states of the matter can be disentangled but this is not the case. Indeed, the ground state of the toric code (or other topologically ordered systems), cannot be disentangled by adiabatic evolution with any local Hamiltonian\cite{nontrivial}.

\section{Topological order away from equilibrium}

\begin{figure}\label{quenchnum}
  \includegraphics[height=.25\textheight]{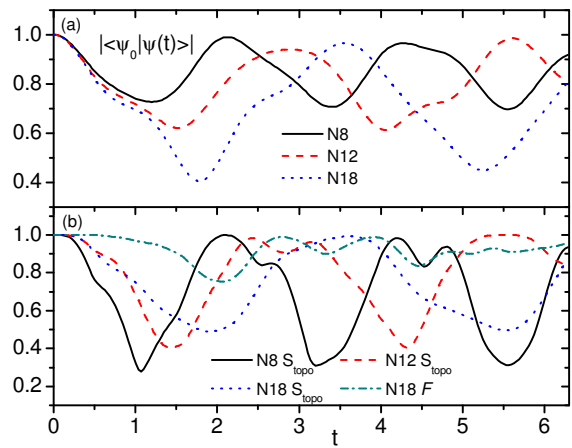}
  \caption{(a) Fidelity of the state at time $t$ with the initial state after the quantum quench with a uniform field. The black line is for a system with only $8$ spins, while the red line is for a $12$ spins system, and the blue dotted one for $18$ spins. This plot shows that as system size is increased, fidelity is decreased and recurrence times are quickly increasing. (b) Here we show the topological entropy (with $\alpha=1$) for the same system sizes. We see that as the system size increases, the topological entropy stays on average closer to the initial value.}
\end{figure}
Now that we know that topological order corresponds to a nontrivial pattern of entanglement present in a wave function, and that we can measure it by the topological R\'enyi entropy, we can ask whether any wave function possesses topological order or not, regardless this state being the equilibrium state of a system, like the ground state or a general Gibbs state at finite temperature. We can therefore ask whether a physical system has topological order away from equilibrium. We simply compute (or measure) its topological $2-$R\'enyi entropy and check. We like the $\alpha=2$-R\'enyi entropy more because this is observable\cite{abanin}, unlike entanglement entropy.

\begin{figure}\label{gab}
  \includegraphics[height=.16\textheight]{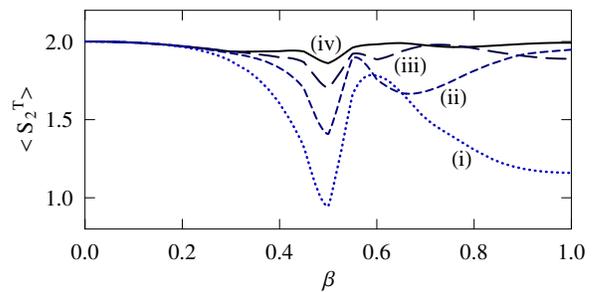}
  \caption{Exact long-term time average of $S_2^T$ as a
function of $\beta = h / (1 + h)$ for various system
sizes: $N = 40$ and $D = 3d = 6$ (i); $N = 60$ and $D = 3d = 9$
(ii); $N = 80$ and $D = 3d = 12$ (iii); $N = 100$ and $D = 3d = 15$
(iv)}
\end{figure}

The scheme we use to cast the system away from equilibrium is the so called {\em quantum quench}. We prepare the system in the ground state of $H_{TC}$ and then, at time $t=0$, we switch on an external magnetic field $h$. The state now will evolve unitarily with the new Hamiltonian $H = H_{TC} -h\sum_l \sigma^z_l$. Notice that such system is non-integrable, and it is therefore expected to thermalize\cite{Polkovnikov}. Also notice that this particular perturbation does not break the gauge structure, which helps in the numerical calculations. Exact diagonalization using Lanczos method allows for the exact knowledge of the evolution operator. The result, originally obtained in\cite{Tsomokos}, shows that as system size increases, whereas the state gets farther from the initial state, its topological entropy is, on average, closer to the initial value, suggesting that in the thermodynamic limit it would be constant, see Fig.\ref{quenchnum}. The main problem with this result, is that, by the method used, we are limited to very small system sizes, up to just $18$ spins. So saying that in the thermodynamic limit something would happen, is at best, an educated guess. We need some better method. Following an intuition first given in \cite{Yu}, we can actually turn the toric code in external field into an integrable system. All we need to do is to apply the external field (in the $z$-direction) only on a subset $H$ of spins (namely, the horizontal ones, black circles in Fig.\ref{lattice}). We can also add another field (in, say $x$ direction) on the spins on the vertical bonds, namely in the subset $V$ (white circles in Fig.\ref{lattice}). This would break the gauge structure, but not integrability\cite{gab1}. The Hamiltonian now reads
\begin{equation}
\hat{H} (\lambda) = - \sum_s \hat{A}_s - \sum_p \hat{B}_p - h
\sum_{i \in H} \hat{\sigma}_i^z - \kappa h \sum_{i \in V}
\hat{\sigma}_i^x, \label{eq-H-1}
\end{equation}
The system is integrable, because it maps onto the array of one dimensional Ising chains\cite{Yu, gab1}, and the exact time evolution can be obtained\cite{gab2}. Computing the $2-$R\'enyi entropy, though, is a much more difficult matter. In order to simplify the calculations, we set $\kappa=0$. Moreover, we can consider just the spins on the boundary of the subsystems of Fig.\ref{Figure2}. This allows to compute the $2-$R\'enyi entropy as a sum of a certain number of correlation functions\cite{gab1}. In \cite{gab2} the time evolution was explicitly computed for a system of $N$ spins. The number of correlation functions to compute grows exponentially with $N$ so this is the only limitation of the method. The results are shown in Fig.\ref{gab}. We can clearly see that as the system size increases, the time average of the topological $2-$R\'enyi entropy $S^{top}_2$ is quickly stabilizing to the initial value. Interestingly, the most destructive quench is to a critical system. 
This result is very good, and reliable, definitely we can now conclude that, for the system considered, topological order is resilient after the quantum quench. 
%\begin{figure}
%  \includegraphics[height=.25\textheight]{Fig_1}
%  \caption{sssss}
%\end{figure}
\section{open problems}
All good then? Well, not so much. There are several open problems, both technical and conceptual. First of all, our results suffer from a big limitation. We have never been able to break integrability and the gauge structure at the same time, simply because then we do not know how to handle the system. It is somehow very surprising that topological order would be resilient if we do both. If we break integrability, thermalization should occur. In some sense a quantum quench would be similar to thermal noise. We expect\cite{chamon,hastingsT} topological entropy to vanish at finite temperature. Here it seems that -from the very limited size numerical calculations performed- even breaking integrability topological entropy is preserved. So either the small size results cannot be trusted, or the gauge structure plays an important role, or there is something about the thermalization of topological order that is defying our understanding. The methods of \cite{gab1,gab2} allow to go to large system sizes, but not to break integrability or gauge structure, although progress can be made in this direction. Indeed, by using the methods in \cite{sid}, one can now compute $S^{top}_2$ for subsystems with a bulk, which is the thing to do if one has to break either integrability or the gauge structure. 

To conclude, topological order is a very exotic and rich playground, with many open questions, especially regarding the behaviour away from equilibrium, the approach to it, and the possible implementations of measurements that can detect it in the lab. All these questions require extensive investigation and are of  importance for theoretical condensed matter and modern quantum statistical mechanics.

%%%%%%%%%%%%%%%%%%%%%%%%%%%%%%%%%%%%%%%%%%%%%%%%
%% BACKMATTER
%%%%%%%%%%%%%%%%%%%%%%%%%%%%%%%%%%%%%%%%%%%%%%%%

\section{theacknowledgments}
This work was supported in part by the National Basic Research Program of China Grant 2011CBA00300, 2011CBA00301 the National Natural Science Foundation of China Grant 61073174, 61033001, 61061130540.

%%%%%%%%%%%%%%%%%%%%%%%%%%%%%%%%%%%%%%%%%%%
%% The following lines show an example how to produce a bibliography
%% without the help of the BibTeX program. This could be used instead
%% of the above.
%%%%%%%%%%%%%%%%%%%%%%%%%%%%%%%%%%%%%%%%%%%

\end{document}